\documentclass[twocolumn,amsmath,amssymb,superscriptaddress]{revtex4-1}

\usepackage{bm}
\usepackage[dvipdfmx]{graphicx}
\usepackage{braket} %Dirac's braket
\usepackage{bbm} %Font
\usepackage{diagbox} % For table
\usepackage{colortbl} % For table
\usepackage{dcolumn}% Align table columns on decimal point

\def\rr{{\bm r}}
\def\RR{{\bm R}}
\def\nn{{\bm n}}
\def\SS{{\bm S}}

\def\tt{{\hat{\bm t}}}

\def\dn{{\delta n}}
\def\bdn{\delta {\bm n}}

\begin{document}

\title{Interactions between atomic-scale skyrmions in 2D chiral magnets}

	\author{Mai Kameda}
	\affiliation{Toyota Central R\&D Labs., Inc., Nagakute, 480-1192, Japan}
	
	\author{Koji Kobayashi}
	%\affiliation{Department of Physics, Kyushu University, Fukuoka 819-0395, Japan}
	\affiliation{Physics Division, Sophia University, Tokyo 102-8554, Japan}

	\author{Yuki Kawaguchi}
	\affiliation{Department of Applied Physics, Nagoya University, Nagoya 464-8603, Japan}	
	\affiliation{Research Center for Crystalline Materials Engineering, Nagoya University, Nagoya 464-8603, Japan}	

\date{\today}

\begin{abstract}
To commercialize stable, highly-integrated, and low-power-consuming magnetic memories of skyrmions, stable control of their motion is crucial. 
Manipulating the inter-skyrmion interactions is key to achieving this goal.
We find that distortion of the skyrmion shape can induce the inter-skyrmion attraction between atomic-scale small skyrmions, similar to larger skyrmions.
Also, the interaction of smaller skyrmions reaches further, when scaled by the skyrmion size, than that of larger skyrmions.
The continuum model is still capable of explaining the behavior.
Moreover, the formation of the magnetic domains between skyrmions also induces attraction, which is again the same mechanism as larger skyrmions.
This study contributes to better control of the skyrmion motion with a wide range of their sizes. 
\end{abstract}
\maketitle

\section{Introduction}
{M}{agnetic skyrmions,} topologically stable magnetic vortices, are potentially capable of carrying information in magnetic memories and computing devices~\cite{NagaosaTokura-nntech, 2016-Finocchio-IOP.49.42.423001, 2017-Fert-Nat.Rev.Mater.2.7.17031, 2018-Everschor-J.Appl.Phys.124.24, 2020-Zhang-IOP32.14.143001, 2021-Tokura-Chem.Rev.121.5.2857}. 
To enhance device controllability, it is important to manipulate inter-skyrmion interactions, which affect the skyrmion motion.
While conventional circular skyrmions have only inter-skyrmion repulsion, 
inter-skyrmion attraction is recently found to appear due to a frustrated exchange coupling~\cite{Rozsa-2016} and due to the three-dimensional magnetic configurations~\cite{Loudon-2018, Du-2018}. 
Recently, we have found two mechanisms of inducing attraction in two-dimensional (2D) chiral magnets~\cite{2021-Kameda-PhysRevB.104.174446}: deformation of the skyrmion shapes and formation of the magnetic domains. The latter can be tuned in two orders of magnitude by tilting the external magnetic field.

For high-density integration of skyrmion magnetic memories, one may utilize recently found small-sized skyrmions, with their radii less than $\sim 3$nm~\cite{Heinze-2011, Romming-2013, Romming-2015, Hanneken-2015, Hsu-2017, 2018-Tomasello-PhysRevB.97.060402, 2019-Hirschberger-Nat.Commun.10.5831, Kurumaji-2019, 2020-Khanh-Nat.Nanotechnol.15.444, 2022-Takagi-Nat.Commun.13.1472}.
Understanding the inter-skyrmion interactions between such small-sized skyrmions, namely atomic-scale skyrmions, is crucial for controlling their motion, though it has not been investigated so far.

In this study, we 
investigate the interactions between such small skyrmions, 
about the size of a few atomic sites {in chiral magnets}. 
We find that the two mechanisms, distortion of the skyrmion
shape and formation of the magnetic domains, are still valid for such small skyrmions. 
As for the attraction due to the distortion of the skyrmion shape, interestingly, the prediction from the analytic expression in the continuum limit is still valid. 
Also, the inter-skyrmion interaction of smaller skyrmions reaches a further distance, scaled by the skyrmion size, than that of larger skyrmions.
To induce the distortion, we either tilt the external magnetic field or use a magneto-crystalline anisotropy.
When the magneto-crystalline anisotropy is sufficiently strong, the magnetic domains are formed between the two skyrmions, which is also the reason for inducing the attraction. These magnetic domains induce strong attraction, which is of the order of the exchange interaction. Remarkably, the magnitude of the attraction remains almost the same regardless of the size of the skyrmions.

The paper is organized as follows. 
In Sec.~\ref{sec:method}, we show the model Hamiltonian
and the method.  
In Sec.~\ref{sec:analytical}, we briefly review the two mechanisms of inducing attractive inter-skyrmion interactions.
In Sec.~\ref{sec:results}, we display numerically investigated inter-skyrmion attractions due to the first mechanism, the deformation of skyrmion shapes. 
In Sec.~\ref{sec:int_Ddependence}, we consider the skyrmion size dependence of the interactions in three setups: under an out-of-plane magnetic field, a tilted magnetic field, and a magneto-crystalline anisotropy.
We compare the numerical results with the approximate potentials in Sec.~\ref{sec:int_compare_Vapp}.
In Sec.~\ref{sec:results2}, we discuss the skyrmion size dependence and comparison with the approximate potentials, of the attraction due to the second mechanism, the formation of magnetic domains.
Finally, we conclude in Sec.~\ref{sec:conclusion}.

\section{Model}%%%%%%%%%%%%%%%%%%%%%%%%%%%%%%%%%%%%%%%%%%%%%%%%%%%%%%%%%%%%%%%%%%%%%%%%%%
\label{sec:method}

The classical spin Hamiltonian of 2D chiral magnets reads
\begin{align}
H=&
\nonumber
-J\sum_{\rr} \SS_{\rr}\cdot (\SS_{\rr + {\bm e}_x}+\SS_{\rr + {\bm e}_y})\\
\nonumber
&-D\sum_{\rr} (\SS_{\rr}\times \SS_{\rr + {\bm e}_x} \cdot {\bm e}_x 
+\SS_{\rr}\times \SS_{\rr + {\bm e}_y} \cdot {\bm e}_y)\\
&-B\sum_{\rr}( S_\rr^z\cos\phi+S_\rr^x\sin\phi)\nonumber\\
&+A\sum_{\rr} \left[(S^x_\rr)^4+\frac{(S^y_\rr+S^z_\rr)^4}{4}+\frac{(-S^y_\rr+S^z_\rr)^4}{4}\right],
\label{eq:aisoHamiltonian}
\end{align}
where $\SS_{\rr}$ is the normalized spin vector on a square lattice $\rr\in\{an_x{\bm e}_x+an_y{\bm e}_y\,|\,n_x,n_y\in\mathbb{Z}\}$ with $a$ being the lattice constant, $J$, $D$, $B$, and $A$ are the coefficients of the exchange interaction, Dzyaloshinskii-Moriya interaction,
Zeeman interaction, and the magneto-crystalline anisotropy, respectively, and $\phi$ is the tilting angle of the magnetic field.
The magneto-crystalline anisotropy in~\eqref{eq:aisoHamiltonian} is the lowest-order potential on cubic lattices that host skyrmions at room temperatures~\cite{Yu-2011, Karube-2016}. 
The anisotropy is on a (011) thin film~\cite{Nagase-2020} that breaks the symmetry of $S^x_\rr$ and $S^y_\rr$, to induce the distortion of skyrmions. 

To investigate the skyrmion-skyrmion interactions, 
we consider two isolated skyrmions excited in a ferromagnetic (FM) state, $\SS_\rr=\tt$. 
This means that we focus on the phase boundary between the FM and the skyrmion crystal phases at low temperatures. 
{The magnitude of the external magnetic field $B=|{\bm B}|$ is close to the second critical field, $B\sim B_{\textrm{cr}2}$}.
The background spins $\SS_\rr=\tt$ are along the resulting easy axes from two competing energies, magneto-crystalline anisotropy and Zeeman energies~\cite{2021-Kameda-PhysRevB.104.174446}. 
{Under an out-of-plane external magnetic field, the background spins are parallel to the external field for $A/B\leq 0.5$. On the other hand, when the magneto-crystalline anisotropy exceeds the critical value $A/B=0.5$, the background spins gradually tilt along the $\pm x$ axis. There are two preferred directions for background spins, $\tt_+$ and $\tt_-$, which are energetically degenerate.}

To numerically calculate the skyrmion-skyrmion interactions, 
we compare the energy of a single skyrmion, $E_\textrm{1sk}$, to that of two skyrmions separated by a relative distance $\RR$, $E_\textrm{2sk}(\RR)$, with respect to the energy of the background uniform configuration $\SS_\rr=\tt$, $E_\textrm{ferro}$. 
The direction $\tt$ is determined by the interplay between the Zeeman energy and the magneto-crystalline anisotropy.
The skyrmion-skyrmion interaction reads
\begin{align}
    V(\RR)=E_\textrm{2sk}(\RR)-2E_\textrm{1sk}+E_\textrm{ferro}.
\end{align}
The energy of each state is obtained by relaxing the Landau--Lifshitz--Gilbert equation at 0{~}K,
$\frac{d\SS_\rr}{dt} = -\SS_\rr\times{\bm B}_{\mathrm{eff}} + \alpha\SS_\rr\times\frac{d\SS_\rr}{dt}$,
where $\bm B_\mathrm{eff}=-\delta H/\delta \SS_\rr$ is the effective magnetic field, $t$ is time, and $\alpha$ is the damping constant.
We pin the positions of skyrmions by a strong local field at each skyrmion core.

\section{Review of two mechanisms for inter-skyrmion attraction}
\label{sec:analytical}

\subsection{First mechanism of attraction -- deformation of skyrmion shapes}

In this section, we derive the approximate expression of the inter-skyrmion interaction in the 
{continuum model by introducing a continuous spin function of $\rr$, $\nn(\rr)$, such that $\nn(\rr)=\SS_\rr$ on the grid points.}
{The approximated interaction well explains the appearance of attraction due to the deformation of skyrmions.}

Suppose that we have a stationary solution of a single-skyrmion state $\nn_{\rm 1sk}(\rr)$,
where a skyrmion at $\rr=\bm 0$ is embedded in a background uniform configuration,
i.e., $\nn_{\rm 1sk}(\bm 0)=-\tt$ and $\nn_{\rm 1sk}(\infty)=\tt$.
The $V(\RR)$ at sufficiently large distance is approximated by~\cite{2021-Kameda-PhysRevB.104.174446}
\begin{align}
{V}_\textrm{app}(\RR)&=\frac{1}{a^2}\int_\Gamma \epsilon_{ij} (A_{-+}-A_{+-})_i dl_j,\\\nonumber
(A_{+-})_i&=\frac{\partial^2f(\tt)}{\partial n_\alpha\partial (\partial_i n_\beta)}\dn_{+,\alpha}\dn_{\rm -,\beta}\\
&+\frac{\partial^2f(\tt)}{\partial (\partial_k n_\alpha) \partial (\partial_i n_\beta)}(\partial_k\dn_{+,\alpha})\dn_{-,\beta},
\label{eq:Vapp}
\end{align}
where $\Gamma$ is the perpendicular bisector, well-distant from two interacting skyrmions, $d\bm l$ is the line element of $\Gamma$ in the direction of ${\bm e}_z\times\RR$, and $\epsilon_{ij}$ is the Levi-Civita symbol, $f(\nn)$ is an energy density of the spin configuration $\nn$, the continuum model of Eq.~\eqref{eq:aisoHamiltonian}, $\nn_+=\nn_{\rm 1sk}(\rr-\RR/2)$, $\nn_-=\nn_{\rm 1sk}(\rr+\RR/2)$, $\nn_{\rm 1sk}$ is single-skyrmion states, and we define $\bdn_\pm\equiv \nn_\pm-(\tt\cdot\nn_\pm)\tt$.
We note that Roman (Greek) indices denote the components in the coordinate (spin) space and take the values $x$ and $y$ ($x, y$, and $z$).
The double-skyrmion state $\nn_{\rm 2sk}$ is obtained from $\nn_+$ and $\nn_-$ using the stereographic projection~\cite{2021-Kameda-PhysRevB.104.174446}.
When we numerically calculate the interactions, $\nn_{\rm 2sk}$ is obtained by placing two skyrmions and relaxing. 
The more detailed derivation is given in {Ref.~\cite{2021-Kameda-PhysRevB.104.174446}.}

The approximate potential {$V_\textrm{app}(\RR)$ of two skyrmions located along the $x$ direction, $\RR=R{\bm e}_x$ and $d{\bm l}\parallel{\bm e}_y$,} under the model~\eqref{eq:aisoHamiltonian} composed of two terms,
\begin{align}
&V_\textrm{app}(R{\bm e}_x)=(J\mathrm{\ term})+(D\mathrm{\ term}),\nonumber\\
&(J\mathrm{\ term})=2J\int_{-\infty}^\infty \left[\partial_x m_x(R/2,y)\right]m_x(-R/2,y) dy\nonumber\\
&\ \ \ \ \ \ \ \ \ \ \ +2J\int_{-\infty}^\infty \left[\partial_x m_y(R/2,y)\right]m_y(-R/2,y) dy,\label{eq:Vint_J}\\
&(D\mathrm{\ term})=\frac{2D}{a} \sin\varphi \cos\chi\int_{-\infty}^\infty dy\nonumber\\
&\left[m_x\left(-\frac{R}{2},y\right)m_y\left(\frac{R}{2},y\right)-m_x\left(\frac{R}{2},y\right)m_y\left(-\frac{R}{2},y\right)\right],
\label{eq:Vint_D}
\end{align}
where $m_x$ and $m_y$ correspond to the components of $\bm n$ projected onto the perpendicular plane to $\tt=(\cos\chi\sin\varphi, \sin\chi\sin\varphi,\cos\varphi)$~\cite{2021-Kameda-PhysRevB.104.174446}. 
The $J$ term is always finite, whereas the $D$ term, {which is proportional to $\sin\varphi$,} becomes finite only when the background spins are tilted from ${\bm e}_z$.
The $J$ term leads to the attraction when the sign of the summation $2J\int_{-\infty}^\infty \sum_{\alpha=x,y}\left[\partial_x m_\alpha(R/2,y)\right]m_\alpha(-R/2,y) dy$ becomes negative. 
{For circular skyrmions, each $\alpha=x,y$ term yields different sign. $\alpha=x$ gives rise to negative and $\alpha=y$ gives rise to positive, and the latter dominates. This balance can be easily flipped by the deformation of the skyrmion shapes, inducing attraction~\cite{2021-Kameda-PhysRevB.104.174446}.}
Distorted skyrmions by the tilted magnetic field and the magneto-crystalline anisotropy in the Eq.~\eqref{eq:aisoHamiltonian} are two examples of inducing such attraction~\cite{2021-Kameda-PhysRevB.104.174446}.
When the background spins are tilted from the $z$ direction, the $D$ term also contributes to the interaction.
However, we find that it does not play a major role in inducing attraction when the size of skyrmions is sufficiently larger than the lattice constant, i.e., $D/J\lesssim0.5$.
This leads to the fact that the influence of the skyrmion shape on the inter-skyrmion interaction for large skyrmions is mostly determined by the $J$ term.

{A question naturally arises about how the inter-skyrmion interaction changes when the contribution from the $D$ term becomes significant. Also, the validity of the $V_\textrm{app}(R)$ for the atomic-scale skyrmions has been unclear.}

\subsection{Second mechanism of attraction -- formation of magnetic domains}%%%%%%%%%%%%%%%%%%%%%%%%%%%%%%%%%%%%%
\label{sec:analytical2}

With strong magneto-crystalline anisotropy $A/B>0.5$, two skyrmions strongly {attract each other} by creating a magnetic domain along $\tt_-$ between them in the background spins along $\tt_+$~\cite{2021-Kameda-PhysRevB.104.174446}. 
The creation of the domain along $\tt_-$ produces an energy profit for the bounded skyrmion pairs, compared to the two isolated skyrmion pairs.

{The strength and range of the attraction are closely related to the area of the domain along ${\tt}_-$. Thus, if the increase of $D/J$ shrinks the length scale with keeping the skyrmion structure, i.e., if the relative number of spins forming the magnetic domains to the size of the skyrmions is preserved for larger $D/J$, the inter-skyrmion potential is expected to be unchanged.}

\section{$V(R)$ due to the shape deformation}%%%%%%%%%%%%%%%%%%
\label{sec:results}
\subsection{Skyrmion--size dependence of $V(R)$}%%%%%%%%%%%%%%%%%%
\label{sec:int_Ddependence}
\begin{figure}[h]
\begin{center}
\includegraphics[width=\linewidth,angle=0]{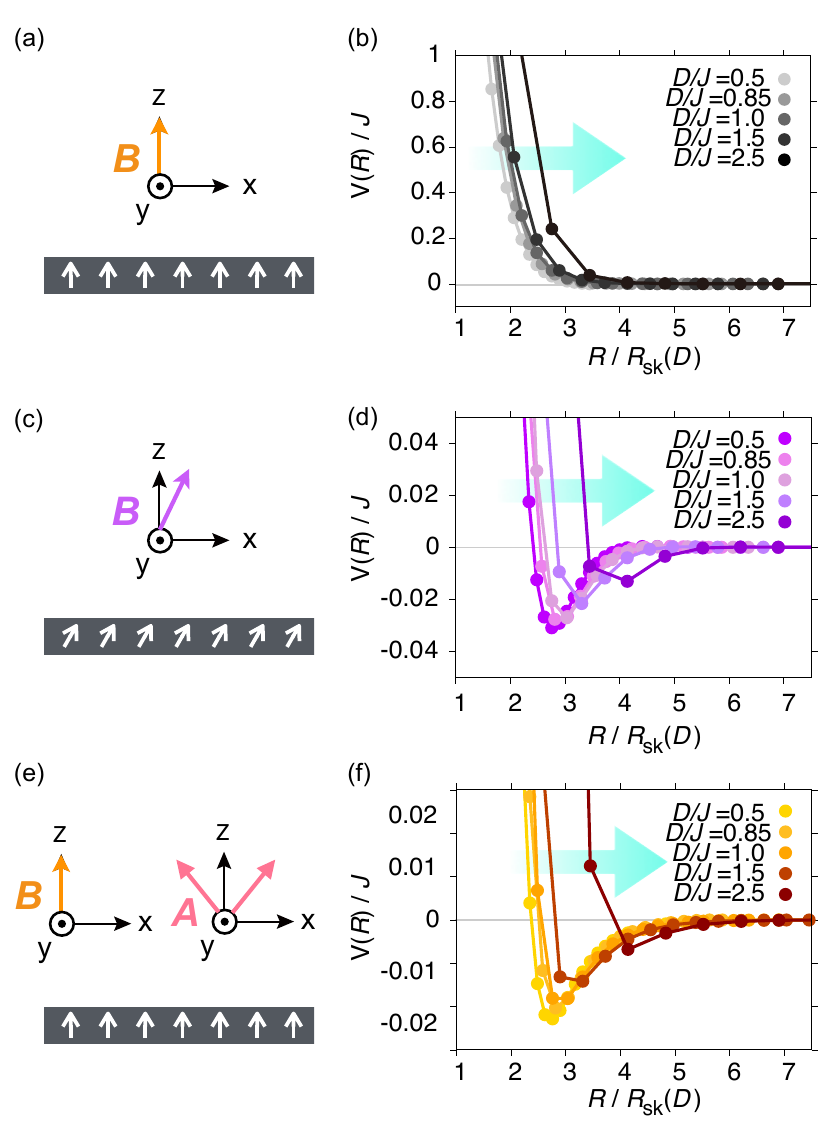}
\end{center}
\caption{
(a),(c),(e) Schematic images of each setup for calculating the inter-skyrmion interactions, under the out-of-plane magnetic field (a), under the tilted magnetic field (c), and under the weak magneto-crystalline anisotropy (e). (b),(d),(f) Inter-skyrmion interactions $V(R)$ of different skyrmion sizes or $D/J$, under the out-of-plane magnetic field of strength $B/D^2=0.75$ (b), under the tilted magnetic field of tilting angle $\phi=30^\circ$ and $B/D^2=0.73$ (d), and under the weak magneto-crystalline anisotropy of anisotropy constant $A/B=0.4$ and $B/D^2=0.7$ (f). 
}
\label{fig:Vr_Dcompare_shift}
\end{figure}
% %%%%%%%%%%%%%%%%%%%%%%%%%%%%%%%%%%%%%%%%%%%%%%%%%
\begin{figure*}[!]
\begin{center}
\includegraphics[width=\linewidth,angle=0]{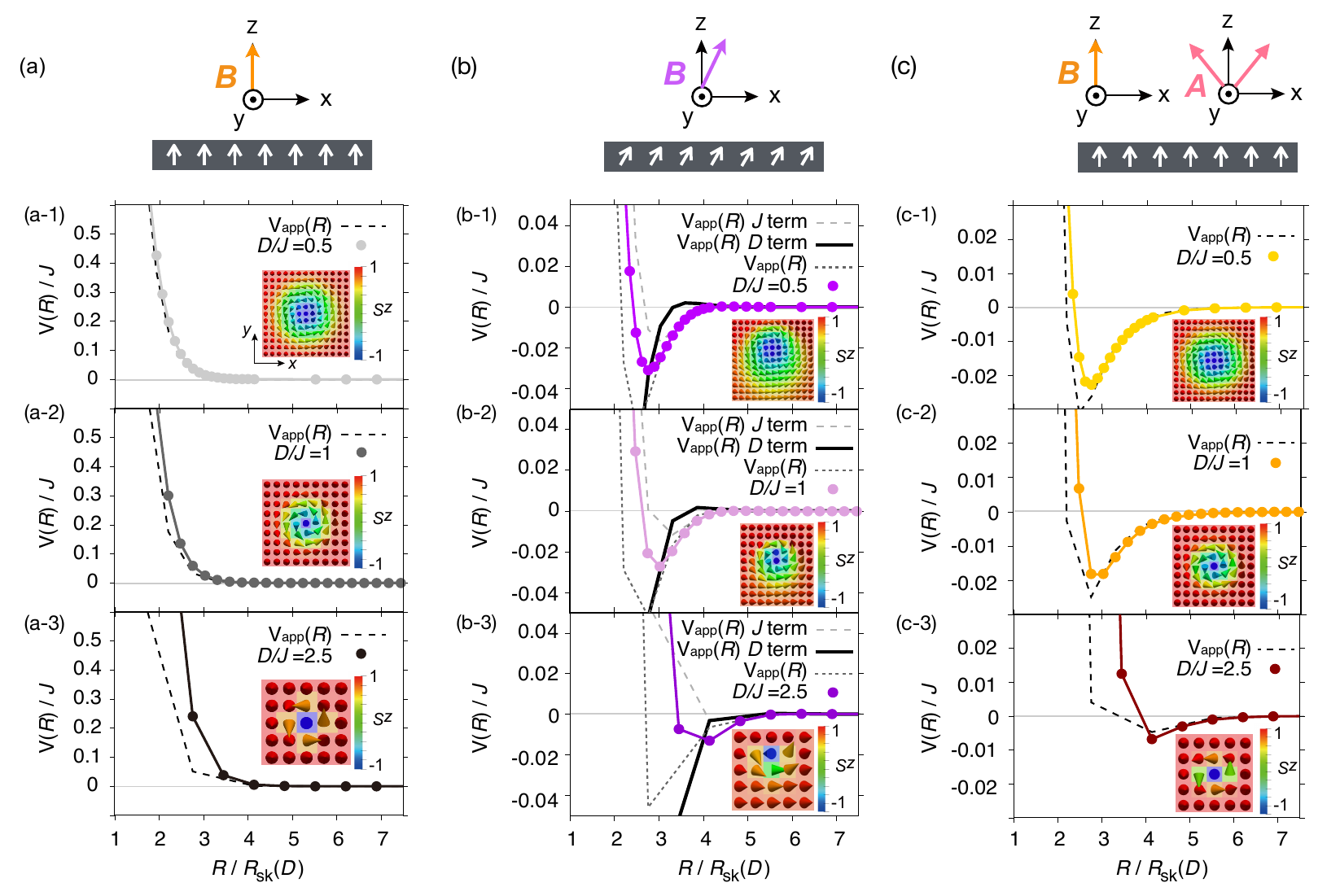}
\end{center}
\caption{
(a)--(c) Schematic images of each setup for calculating the inter-skyrmion interactions. (a-1)--(c-3) Comparisons of numerically obtained interactions $V(R)$ and the approximate potential $V_\textrm{app}(R)$ under the out-of-plane magnetic field with $B/D^2=0.75$ (a-1)--(a-3), under the tilted magnetic field with $\phi=30^\circ$ and $B/D^2=0.73$ (b-1)--(b-3), and under the weak magneto-crystalline anisotropy with $A/B=0.4$ and $B/D^2=0.7$ (c-1)--(c-3). In (b-1)--(b-3), for which Eq.~\eqref{eq:Vint_D} is non-vanishing,  we also show the contributions of the $J$ and $D$ terms to $V_\textrm{app}(R)$. Inset in each panel shows the stable one skyrmion state. The panel sizes of the insets are $(1.8R_{\textrm{sk}})^2$, $(2.5R_{\textrm{sk}})^2$, and $(3.4R_{\textrm{sk}})^2$, for $D/J=0.5$, $D/J=1$, and $D/J=2.5$, respectively.
}
\label{fig:Vr_Vapp_compare}
\end{figure*}
% %%%%%%%%%%%%%%%%%%%%%%%%%%%%%%%%%%%%%%%%%%%%%%%%%

Here we explore the skyrmion size dependence of the interactions, with changing $D/J$ in the range of $0.5\leq D/J\leq 2.5$.
We investigate three setups, under an out-of-plane magnetic field, under a tilted magnetic field, and under a magneto-crystalline anisotropy. 
In all cases, we find that the inter-skyrmion interaction of smaller skyrmions reaches further distance with respect to the skyrmion size than that of larger skyrmions.

In Figure~\ref{fig:Vr_Dcompare_shift}, we show the skyrmion-skyrmion pair potential along the $x$ direction $V(R\bm e_x)\equiv V(R)$ for various values of $D/J$. 
The distance $R$ is scaled by the radius of skyrmions $R_{\mathrm{sk}}$.
We use the radius $R_{\mathrm{sk}}=\frac{2\pi J}{\sqrt{3}D}$ from the continuum model~\cite{NagaosaTokura-nntech, kawaguchi-2016}, which is found to be a good approximation of the skyrmion size even with large $D/J$; 
the estimation of skyrmion size with polynomial interpolation, from numerically obtained spin configurations, turns out very similar to the continuum $R_{\mathrm{sk}}$.

When we increase $D/J$ under an out-of-plane field [Fig.~\ref{fig:Vr_Dcompare_shift}(a)], the repulsion reaches further with respect to the skyrmion size $\sim R_{\mathrm{sk}}$ [Fig.~\ref{fig:Vr_Dcompare_shift}(b)].
{This is because a skyrmion becomes stiffer against shrinkage as its size becomes closer to the lattice constant.}
The circular skyrmion does not induce attractive inter-skyrmion interaction.

An in-plane magnetic field with angle $\phi=30^{\circ}$ [Fig.~\ref{fig:Vr_Dcompare_shift}(c)] 
and the weak magneto-crystalline anisotropy $A/B=0.4\le 0.5$ [Fig.~\ref{fig:Vr_Dcompare_shift}(e)], 
produces the attraction along the $x$ direction.
{In both cases, the hard-core repulsion starts at larger $R/R_{\textrm{sk}}$ as $D/J$ increases as in the case of Fig~\ref{fig:Vr_Dcompare_shift}(a).}
The emergent attraction reaches further with respect to the skyrmion size $\sim R_{\mathrm{sk}}$ for larger $D/J$ [Fig.~\ref{fig:Vr_Dcompare_shift}(d) and (f)].

\subsection{Comparison of $V(R)$ and $V_\textrm{app}(R)$}%%%%%%%%%%%%%%%%
\label{sec:int_compare_Vapp}

{To understand the behavior of $V(R)$ at a distance, we compare it with $V_\textrm{app}(R)$ in Eq.~\eqref{eq:Vint_J} and Eq.~\eqref{eq:Vint_D}.}
Though the approximate potential $V_\textrm{app}(R)$ is based on the continuum model, we find that it well agrees with $V(R)$ even for smaller skyrmions; 
the spatial derivative of spins become{s} small enough at sufficiently large $R$.

The comparison of $V(R)$ and $V_\textrm{app}(R)$ are shown in Fig.~\ref{fig:Vr_Vapp_compare} for three different setups. 
In all cases of Fig.~\ref{fig:Vr_Vapp_compare}, $V(R)$ and $V_\textrm{app}(R)$ agree well even at larger $D/J$, for sufficiently large $R$;
{Figs.~\ref{fig:Vr_Vapp_compare}(a-3), (b-3){,} and (c-3) show} that $V(R)$ is reproduced by the $V_\textrm{app}(R)$ at $R/R_{\textrm{sk}}\gtrsim 4.7$, where $V(R)$ is nonzero.
$V_\textrm{app}(R)$ is calculated from numerically obtained stable skyrmion shapes, which shrink with keeping the in-plane structure [Inset of Figs.~\ref{fig:Vr_Vapp_compare}(a-1)--(a-3), (b-1)--(b-3), (c-1)--(c-3)].

Under the tilted magnetic field [Fig.~\ref{fig:Vr_Vapp_compare}(b)], 
the background spins are tilted and the $D$ term of $V_\textrm{app}(R)$, Eq.~\eqref{eq:Vint_D}, becomes finite. The $D$ term contributes dominantly to the attraction and reaches further $R/R_{\mathrm{sk}}$ as $D/J$ increases, giving rise to potential minimum at further $R/R_{\mathrm{sk}}$ [Fig.~\ref{fig:Vr_Vapp_compare}(b-1)--(b-3)].

Under the weak magneto-crystalline anisotropy $A/B \leq 0.5$, $A/B=0.4$ [Fig.~\ref{fig:Vr_Vapp_compare}(c)],
the background spins are not tilted and the $D$ term Eq.~\eqref{eq:Vint_D} is zero.
The $J$ term of $V_\textrm{app}(R)$ decides the inter-skyrmion interaction, which may reflect the small change of the skyrmion shape due to larger $D/J$.
$V_\textrm{app}(R)$ well reproduces $V(R)$ at large $R$ [Fig.~\ref{fig:Vr_Vapp_compare}(c-1)--(c-3)].
Larger $D/J$ merely changes $V(R)$ though the attraction reaches a bit further $R/R_{\textrm{sk}}$  [Fig.~\ref{fig:Vr_Vapp_compare}(c-3)].

% %%%%%%%%%%%%%%%%%%%%%%%%%%%%%%%%%%%%%%%%%%%%%%%%%
\begin{figure}[!]
\begin{center}
\includegraphics[width=\linewidth,angle=0]{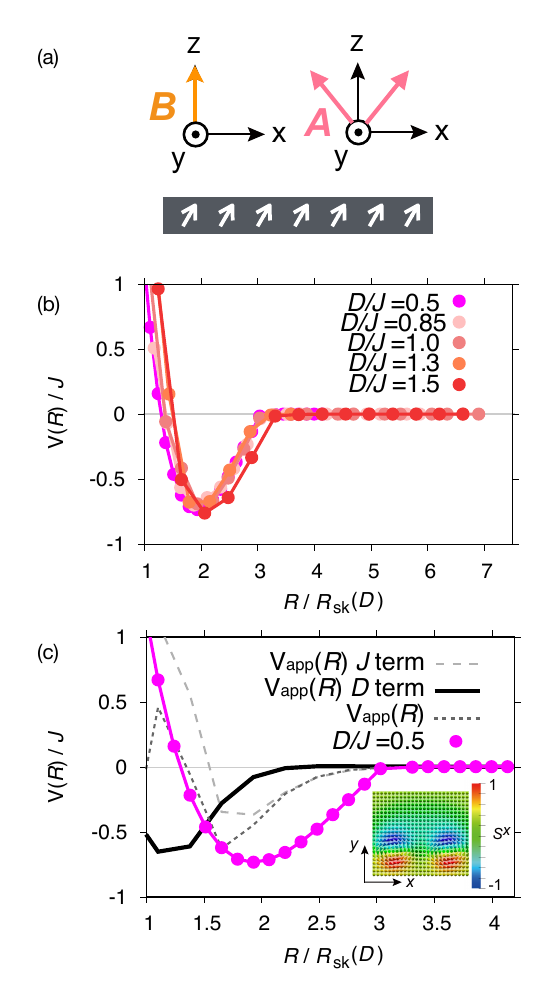}
\end{center}
\caption{
(a) Schematic image of the setup for calculating the inter-skyrmion interactions of skyrmions with magnetic domains. (b) The $D/J$ dependence of the inter-skyrmion interactions $V(R)$ with $A/B=1.0$ and $B/D^2=0.7$. The solid lines are a guide for the eye. (c) Comparison of $V_\textrm{app}(R)$ and $V(R)$ with $D/J=0.5$. Other parameters are the same as that of (b).
The dashed line shows the $J$ term, the thick solid line shows the $D$ term, and the dotted line shows the total $V_\textrm{app}(R)$. The circles show the $V(R)$. Inset shows the stable domain wall skyrmion state. The panel sizes of the inset is $3.0R_{\textrm{sk}}\times3.4R_{\textrm{sk}}$.}
\label{fig:Vr_largeD_DWSk}
\end{figure}
% %%%%%%%%%%%%%%%%%%%%%%%%%%%%%%%%%%%%%%%%%%%%%%%%%

\section{$V(R)$ due to the domain formation}%%%%%%%%%%%%%%%%%%
\label{sec:results2}

We explore the skyrmion size dependence of the interactions, with changing $D/J$ in the range of $0.5\leq D/J\leq 1.5$.
We note that exploring $D/J>1.5$ was impossible in this case, since the skyrmions become too small to be recognized.
With strong magneto-crystalline anisotropy $A/B>0.5$ [Fig.~\ref{fig:Vr_largeD_DWSk}(a)], two skyrmions embedded in a FM state $\tt_+$ are strongly bounded by creating a magnetic domain along $\tt_-$ in-between [Inset of Fig.~\ref{fig:Vr_largeD_DWSk}(c)].

Remarkably, large $D/J$ merely changes the potential depth and distance $R/R_{\mathrm{sk}}$ which gives the potential minimum. Also, attraction reaches a similar range [Fig.~\ref{fig:Vr_largeD_DWSk}(b)].
{Enhancement of $D/J$ may affect the number of spins in the magnetic domain along $\tt_-$ and on the domain wall, though the value of $D/J$ does not affect the total energy inside the magnetic domain along $\tt_-$.}
This suggests that the strong inter-skyrmion attraction would be observed
regardless of the skyrmion size.
Larger $D/J$ again just shrinks the spin configuration (not shown).
Due to the formation of the magnetic domain, the attraction becomes much stronger than predicted $V_\textrm{app}(R)$, of the order of $\sim J$ [Fig.~\ref{fig:Vr_largeD_DWSk}(c)].
The disagreement between $V(R)$ and $V_\textrm{app}(R)$ may be attributed to the different origins of these attractions, as discussed in {Ref.~\cite{2021-Kameda-PhysRevB.104.174446}.}

\section{Conclusion}
\label{sec:conclusion}
We find that two mechanisms for producing attractive inter-skyrmion interactions, the deformation of the skyrmion shapes and the formation of the magnetic domains, are also valid for the atomic-scale skyrmions. 
When the skyrmion gets smaller, the order of the magnitude of attractions merely changes; the former mechanism induces weak attraction, of the order of $\sim 0.01J$, and 
the latter induces strong attraction, as much as $\sim J$. 
{Meanwhile,} the skyrmion-skyrmion pair potential reaches further distance $R/R_{\textrm{sk}}$.
These findings may accelerate further research to utilize skyrmions as information carriers.
Our results would give a guiding principle for manipulating inter-skyrmion interactions in more realistic setups.

\section{Acknowledgement}
This work was supported by JSPS KAKENHI (Grants No.~JP21H01009 and No.~JP22K03446).

\bibliography{refs}
\end{document}